\documentclass[aps,prb,reprint,superscriptaddress]{revtex4-1}
\usepackage{etoolbox}
\usepackage{graphicx,amssymb,bm,amsfonts,amsmath,graphics,color,epstopdf}
\usepackage[bookmarks=false,pdfstartview=FitH,colorlinks=true,citecolor=blue,linkcolor=blue]{hyperref}
\usepackage[version=3]{mhchem}
\usepackage{lettrine}

\begin{document}

\title{Reflections on the Physics and Astronomy \\ Student Reading Society (PhASRS) at San Jos\'e State University}

\author{Sidney L.\ Johnson}
\affiliation{Department of Physics and Astronomy, San Jos\'e State University, San Jose, CA 95192, USA}
\author{Athanasios  Hatzikoutelis}
\affiliation{Department of Physics and Astronomy, San Jos\'e State University, San Jose, CA 95192, USA}
\author{Christopher L.\ Smallwood}
\email[Email:\ ]{christopher.smallwood@sjsu.edu}
\affiliation{Department of Physics and Astronomy, San Jos\'e State University, San Jose, CA 95192, USA}
\date {\today}

\maketitle

\lettrine[findent=2pt]{\textbf{T}}{}he COVID-19 pandemic imposed profound changes on the way we think about undergraduate physics education. Online courses became mainstream. Exam formats were reimagined. Digital whiteboards replaced face-to-face discussions. Laboratory classes were outfitted with home-delivered supply kits. And all of us developed a more intimate knowledge of Greek letters and symbols (delta, omicron, etc.) than we might have comfortably liked to admit.
 
Having weathered these transformations from the point of view of both an undergraduate student (S.L.J.) and classroom instructors (A.H. and C.L.S.), we have found that among the most challenging aspects of the in-person learning experience to replicate in an online environment have been the relational ones. To highlight some of the ways in which these issues can be mitigated, we report here on the activities of the San Jos\'e State University (SJSU) Physics and Astronomy Student Reading Society (PhASRS), which was an online reading group at SJSU founded by ourselves and others running from the summer of 2020 until the end of the fall 2020 semester. Elements of the reading group's structure and guiding principles are described, as well as student and faculty reflections on what worked well and what did not. The manuscript underlines the power of astronomy- and physics-themed journal clubs as vehicles for learning\cite{DiTusa2006,Dake2018,Cetnar2021,Santos2021} and more generally emphasizes the importance of community-building initiatives in the discipline.\cite{Corbo2013,Sabella2017,Quan2019,Brown2021,Rethman2021} Our hope is that this summary of activities will inspire faculty members and students at colleges and perhaps high schools to imagine new possibilities for developing communities of people in science that might not otherwise be able to exist. 

\begin{figure}[tb]
	\centering
	\includegraphics[width=3.375in]{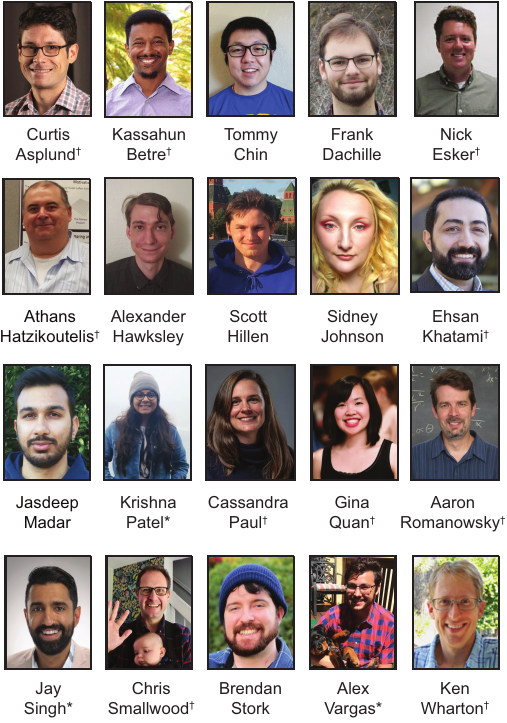}
	\caption{\label{yearbook}A sampling of San Jos\'e State University PhASRS participants. Names with asterisks (*) indicate master's students. Names with daggers ($\dagger$) indicate faculty and staff. All other participants were undergraduates.}
\end{figure} 

\section{Origins and goals}

The idea for the PhASRS reading group arose from multiple directions simultaneously. Among the primary motivators was the fact that in the summer of 2020, during the height of the pandemic, internship and research programs that would usually have been conducted over the summer were largely canceled. ``In the beginning of the pandemic, I had nothing to do," said Frank Dachille, a physics major and college junior at the time of the program. ``I really wanted to do some physics." This sentiment and others like it were communicated in class to one of us (C.L.S.) toward the end of the spring 2020 semester, and discussions were initiated at an end-of-year departmental faculty meeting about what could be organized in response. In July of 2020 the reading group was born with Smallwood as lead facilitator, and the group featured nine students and three faculty members and lecture staff at its inaugural meeting.

\begin{figure*}[tb]
	\centering
	\includegraphics[width=7in]{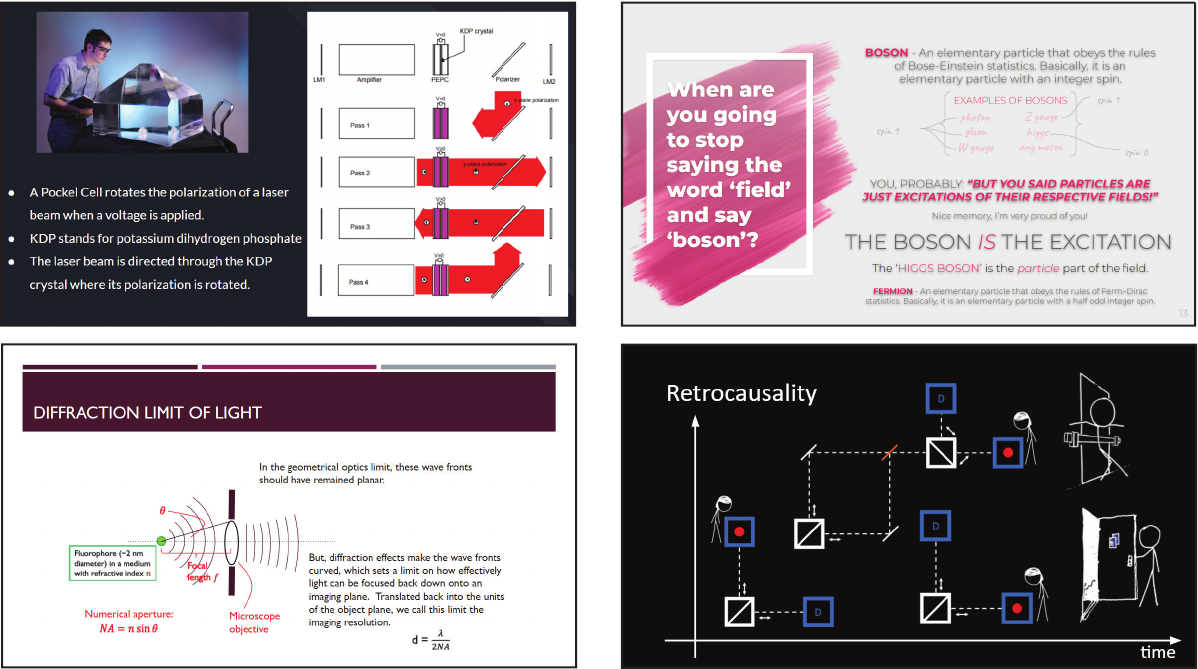}
	\caption{\label{slides}Slide samples from the SJSU PhASRS program. Images courtesy of Frank Dachille (upper left), Sidney Johnson (upper right), Jasdeep Madar (lower left), and Tommy Chin (lower right).}
\end{figure*}

Independently of this, another one of us (A.H.) had been exploring the possibility of expanding upon a directed reading course that had been offered during the fall of 2019 aimed at exposing students to the world and career pathways associated with particle physics. Conceived innovations included an expansion of topics into other fields of research to help students better understand the ways in which a physics background can be leveraged. Hatzikoutelis took over the role of facilitating PhASRS as the fall 2020 semester commenced. The group was codified into a one-unit class employing a credit/no credit grading scheme, and the exposure-to-research goal was set as an underlying value. Learning outcomes included the ability to identify and reflect on the salient points of scientific articles.

Among the core values of the reading group for both the summer and academic-year settings was that it be student-centered but also reliant on the involvement of faculty members, serving, in some ways, as a replacement research experience for students who would have otherwise landed internships and as a gateway to research for students just starting to get involved. The sense of shared investment in time and resources remained an important feature of the reading group throughout its existence. For any given reading group session there were 3--5 faculty members attending and 6--10 students (see Fig.~\ref{yearbook}). Nine different faculty members and 12 students participated in the group in total over the combined summer and fall sessions.

\section{Activity Structure}

Structural elements of the reading group can be roughly organized into three categories: meeting format, subject matter, and technological support. 
 
{\bf Meeting format:} PhASRS meetings occurred on a weekly basis, were restricted to one hour, and were conducted over the video chat platform Zoom. In recognition that online platforms pose special challenges in terms of fostering community (as opposed to fully in-person formats), an emphasis was placed on relationship-building, with the first several minutes of each reading group session devoted to personal check-ins and informal chat. The first meeting was devoted entirely to getting students and faculty members acquainted with each other, to working out logistical details, and to establishing the reading group name. The second and third meetings were led by faculty members to set examples and templates for student-hosted meetings yet to come (Smallwood led the second meeting with a discussion of an article related to quantum computation, and Dr.\ Ehsan Khatami led the third meeting with a discussion of an article on the Ising Model). Following this, the role of discussion group leader rotated on a weekly basis, with different students or faculty members assembling slide presentations on reading group material each week. In weeks featuring student presenters, the student in charge was paired with a faculty mentor in an arrangement designed to give students a taste of what a one-on-one research experience with the faculty mentor might look like, to strengthen student-faculty relationships, and to provide resources and guidance to students as they prepared their facilitator notes. With the weekly discussion leaders rotating in and out, we found it helpful to also have a lead facilitator charged with long-term logistics, speaker coordination, and student-faculty pairing. As mentioned above, Smallwood filled this role throughout the summer and Hatzikoutelis took over the role in the fall.

Discussion leaders were encouraged to prepare slide sets to go along with the meetings that they facilitated, as exemplified in Fig.~\ref{slides}.  The reasoning behind this was to make participation barriers for the rest of the members of the group as low as possible so that students who might not have had a chance to read the assigned article or who didn't fully understand it would still be able to contribute. That being noted, it can be tempting turn slide sets of this sort into extended monologues conducted exclusively by the facilitator. To help avoid this pitfall, interruptions were welcomed, and facilitators were encouraged to pepper their slides with discussion questions aimed at inviting reactions and thoughts from the rest of the group. Typical goals for the ratio of presentation time to discussion time were about half and half, with facilitators fluidly switching back and forth between these two kinds of interaction modes over the course of the hour. In the liveliest and (to our minds) most successful sessions, questions posed by the facilitator naturally evolved into whole group discussions in which many different people chimed in, and in which the facilitator only occasionally stepped in to steer the conversation forward.
 
{\bf Subject matter:} Reading group subject matter was largely chosen from popular science articles published in venues like {\it Physics Today}, {\it Scientific American}, and the ``News and Views" Section of {\it Nature}. The purpose of selecting articles at this level was to introduce students to modern topics of interest in physics while still keeping content maximally accessible. Depending on the topic, supplementary references to original peer-reviewed academic journal articles were occasionally recommended as well. A list of discussion topics is summarized in Table \ref{table1}, and references are collected in full at the end of this article. Topics were suggested and selected by both faculty members and students. Once a topic was identified, article assignments to go along with it were selected by faculty members (in the case of faculty-led discussions) or by students and their assigned faculty mentors in a collaborative approach (in the case of student-led discussions). In this latter case, the article assignment often grew out of a brief literature review conducted by the student independently or with assistance of the faculty partner.

\begin{table}[tb]
\centering
\caption{Schedule of PhASRS Discussion Topics (Starred discussions were led by students)}
\begin{ruledtabular}
\begin{tabular}{l l r}
Week & Topic & Refs.\ \\
\hline	
1 & Introduction & -- \\
2 & Quantum Computation & [\!\!\citenum{Mermin2000}] \\
3 & The Ising Model & [\!\!\citenum{Wood2020}] \\
4 & The Nature of Quantum Measurement* & [\!\!\citenum{Weatherall2017}] \\
5 & Physics Education Research & [\!\!\citenum{Redish1999}] \\
6 & High-Temperature Superconductivity* & [\!\!\citenum{Castelvecchi2019}] \\
7 & Discovering the Higgs Boson* & [\!\!\citenum{Gray2018}] \\
8 & Scientific Fraud* & [\!\!\citenum{Levi2002}] \\
9 & Superresolution Microscopy* & [\!\!\citenum{Couzin2006,Betzig2006}] \\
10 & Highlights of the SLAC Summer School & [\!\!\citenum{slac}] \\
11 & Variations on Schrodinger's Cat Paradox* & [\!\!\citenum{Merali2020}]  \\
12 & Imaging a Black Hole* & [\!\!\citenum{Castelvecchi2019a}] \\
13 & Recoil Mass Spectrometry & [\!\!\citenum{Schirber2018}] \\
14 & Social Equity Impact of Physics Grade Scales & [\!\!\citenum{Paul2018,Guskey2013}] \\
15 & Neural Networks as Physics Analysis Tools & [\!\!\citenum{Savitsky2020,Aurisano2016}] \\
16 & The Michelson-Morley Experiment* & [\!\!\citenum{Michelson1887}] \\
17 & The Physics of Climate Change* & [\!\!\citenum{climate1}] \\
18 & Fusion and the National Ignition Facility* & [\!\!\citenum{nif}] \\
\end{tabular}
\end{ruledtabular}
\label{table1}
\end{table}
 
{\bf Technological support:} Beyond the weekly video chat meeting held on Zoom, the PhASRS group schedule was maintained through a shared Google Document listing dates, topics, presenters, and (where applicable) faculty mentors. A channel devoted to the PhASRS reading group was also created on a Discord server maintained by the SJSU Physics and Astronomy Club, and students were encouraged to share their reactions to the assigned reading group articles asynchronously in advance of the weekly video chat meeting. While reading the article, a few students assembled detailed article notes in privately created Google Documents, and they shared these notes with the rest of the group so that other students and faculty members could comment and react. For the course-credit version of the reading group in the fall, there was a course page, created on the SJSU-managed learning management system Canvas, with the reading material and student and faculty presentations. There were also discussion assignments in the course-credit version of the group where the students commented on the readings and the presentations for each subject. The attendance and the discussion participation defined the student grades.

\section{What worked? And What didn't?}

To help understand the impacts of PhASRS better, we collected a series of interviews with PhASRS participants in the fall of 2021, a year after the program ended. Interviews were conducted by S.L.J.\ in one-on-one recorded video conversations over Zoom. Student participants reported in these interview sessions that the experience left them more academically prepared for future careers than they otherwise would have been, and that it also brought them closer to their classmates.

Though at first intimidating to students, the hour-long structure of the talks provided many students with important expertise in giving longer presentations, as well as being leaders of discussions. ``It was daunting!" said Tommy Chin, a student who joined PhASRS at the beginning of the Fall 2020 semester as an undergraduate junior. ``To actually fill up the entire hour was daunting. I'd never actually given an hour talk before." Students quickly rose to the challenge, however, with success attributed in some cases to the designation of faculty mentors. ``Pairing us with a mentor really helped," said Chin. ``I don't think I would have been able to do it if it hadn't been  for Dr.\ Wharton." In retrospect, some of the students remarked that giving a longer talk was actually much easier than the usual 10--15 minute talks they give for final projects in school, and in almost all cases, students reported being thankful for the skills developed. ``Now that I have done those two really long talks, it seems like the longer talks are easier," said Frank Dachille. ``It was amazing practice," said Jay Singh, who was a recently graduated SJSU Business Administration major at the time he joined the group, and has since matriculated into the SJSU Physics master's program. Faculty who were PhASRS mentors agreed with the students, both in that longer talks were generally easier, and that both hour-long and shorter presentations were great practice for the students' futures. Dr.\ Ehsan Khatami, one of the PhASRS group faculty mentors, noted that the talk provided a particularly nice practice venue for his master's-degree research student. ``Now that he's graduating, he {\it has} to give a talk like that," Khatami said, ``and he can even maybe recycle a few [of the PhASRS presentation] slides."

Students reported that one of the other notable academic benefits of participating in PhASRS was in learning how to understand and process the scientific content. Scott Hillen, who was an undergraduate junior at the time he joined PhASRS, remarked that one of the things he appreciated most about the group was the way it enabled him to see how professors interact with each other in real-time discussions. ``What drew me in," he said, ``was the ability to see how professors operated, how they looked at, and read, and understood papers. And so basically what would happen is they would present, and I would just be sitting there in awe, and I would think `Oh my God, they think \emph{that}? They're attacking it \emph{this} way?'" The reflections of other students were similar. Most cited that being able not only to give talks, but also to watch ones given by peers and professors gave them the ability to learn and improve their own expertise.  ``When the students presented, it kind of gave all of us an understanding of `hey, this is how we could present, and maybe do better here or there,'" said Singh, continuing, ``When the professors presented, it gave a template of how you should." Chin reflected that ``it reinforced the concepts I was learning in class." 

The program's greatest benefits, however, may have had less to do with the development of individual skills than with the cohort and community that the creation of the reading group helped foster. Students commented that this development was largely facilitated by the group's relaxed atmosphere and emphasis on student leadership. ``The main selling point of the program is that it was an easy-going, non-stressful environment to learn," said Chin. ``And also, it was kind of like a potluck. Everyone brought different things [to discuss]." In consequence, said Chin, students in the PhASRS reading group felt especially free to ask questions. ``It was the little things, right? You're always constantly afraid [in more formal contexts] that your questions are stupid, and people might think that you're asking a stupid question. But that environment made it less intimidating." Alexander Hawksley, an undergraduate junior who joined PhASRS in the fall of 2020 at the same time as Chin, echoed this sentiment. ``It was really nice to have a presentation on very complicated material from someone who did not understand it that well. And that sounds like it would be bad! But it was nice because they knew the presentation was upcoming, it was a relaxed environment so they knew they wouldn't be severely penalized if they didn't have all the answers, and it was okay to say, `I don't know, I'm not an expert.'" Hawksley went on to mention that the course being voluntary and not required also contributed to the environment and his initial interest: ``I knew everybody in the audience wanted to be there, because it was basically a volunteer course," he said. Others mentioned the lack of grades or tests was also a reason for them feeling much less stress than a normal class: ``We all understood that we were there to learn and to gain experience, not to judge anyone else or get grades," said Dachille.

An idea that many students commented was intertwined with the group's relaxed atmosphere was the fact that the same set of students kept on choosing to come back to the group on a recurring basis. ``It makes you realize you're not alone in the boat...There are a lot of people in your situation, or a similar situation, so you don't feel like you're going through it by yourself," said Singh, highlighting the struggle of isolation many students tend to feel at some point in their undergraduate career in physics. ``It built a cohort," said Chin. ``People who were in PhASRS, we ended up taking classes together after that." Students who were involved in PhASRS and went on to take classes together had a tendency towards building study groups outside of class (online at first, of course, due to the pandemic). ``It feels like this community that has free-flowing ideas, and it's cool to see how they form, and how they turn out later." said Dachille. Students were very emphatic about the importance of the culture of PhASRS, which encouraged students to learn and ask questions. Many made it clear that this was different from being in a normal classroom setting, and felt like they were included and encouraged to ask questions and learn from others.

Alongside the positive feedback, students did report a few downsides, with compliments of the program being interwoven on occasion with commentary about the required time commitment. Dr.\ Kenneth Wharton, one of the group's faculty mentors, noted, for example, that ``carving an hour out of your day is challenging sometimes, but people liked it.” Dachille said in his overall assessment of the reading group that ``Time permitting, I would do it again, definitely. I think it's great." Students who participated in PhASRS as it wrapped into the fall noticed the time crunch more acutely than those who participated in the summer. Chin commented on the stress associated with putting his presentation together as midterms for other classes loomed in the fall. ``I remember it was really crunched. I was doing it at the last minute, because PhASRS is kind of like an `extra' on top of school." This lack of time was in fact the biggest reason that students gave that would stop them from participating again in the program. Some students were taking more or harder upper-division courses, and others were concerned about having to apply to graduate school in the upcoming semester.

\section{Summary and Outlook}

In spite of the time-commitment challenges, the impact on students who participated in the program seems to have been deep. It certainly energized us as the authors of this manuscript reflecting back on the experience. PhASRS was created during the beginning of what we didn't know then would be nearly two years of isolation. It was created because students and teachers still wanted to learn together, despite the uncontrollable circumstances. Our impressions are that in many ways we succeeded in fostering this learning.

For the student author among us (S.L.J.), the club acted as a genesis for the author to begin to see herself as a scientist, to be able to understand scientific papers, and to be able to communicate scientific knowledge. Being surrounded by other like-minded students and professors made it easy to feel accepted and fostered a sense of individual confidence. The variety of topics each week made coming to the club exciting, and was a refreshing departure from the usual curriculum-based studying.

For the faculty authors among us (A.H.\ and C.L.S.), the experience provided a potent illustration of student potential and talent. A.H.\ came away from the experience impressed with the ability of the PhASRS meeting format to be used as a vehicle for demonstrating that the world of physics research is far bigger than the typical undergraduate physics curriculum conveys. C.L.S.\ found himself amazed by the level of creativity and leadership that emerged from PhASRS students as they led group meetings, and he was struck by the durability of the relationships that the reading group helped form.

As we write up this summary, many of the students who participated in PhASRS have just recently graduated from SJSU and are beginning to enter the world beyond. A significant fraction of students have elected to pursue graduate studies in physics. Others have sought and obtained positions in private industry. It will be exciting to see where these different career paths lead, and our hope is that the PhASRS program helped---in at least a small way---to illuminate possibilities.

\begin{acknowledgments}
We thank G.\ M.\ Quan and M.\ E.\ Kress for useful discussions, and the many faculty and student PhASRS participants for sharing their time and perspectives in one-on-one interviews. 
This material is based upon work supported by the National Science Foundation under Grant No.\ 2003493.
\end{acknowledgments}


\begin{thebibliography}{29}%
\makeatletter
\providecommand \@ifxundefined [1]{%
 \@ifx{#1\undefined}
}%
\providecommand \@ifnum [1]{%
 \ifnum #1\expandafter \@firstoftwo
 \else \expandafter \@secondoftwo
 \fi
}%
\providecommand \@ifx [1]{%
 \ifx #1\expandafter \@firstoftwo
 \else \expandafter \@secondoftwo
 \fi
}%
\providecommand \natexlab [1]{#1}%
\providecommand \enquote  [1]{``#1''}%
\providecommand \bibnamefont  [1]{#1}%
\providecommand \bibfnamefont [1]{#1}%
\providecommand \citenamefont [1]{#1}%
\providecommand \href@noop [0]{\@secondoftwo}%
\providecommand \href [0]{\begingroup \@sanitize@url \@href}%
\providecommand \@href[1]{\@@startlink{#1}\@@href}%
\providecommand \@@href[1]{\endgroup#1\@@endlink}%
\providecommand \@sanitize@url [0]{\catcode `\\12\catcode `\$12\catcode
  `\&12\catcode `\#12\catcode `\^12\catcode `\_12\catcode `\%12\relax}%
\providecommand \@@startlink[1]{}%
\providecommand \@@endlink[0]{}%
\providecommand \url  [0]{\begingroup\@sanitize@url \@url }%
\providecommand \@url [1]{\endgroup\@href {#1}{\urlprefix }}%
\providecommand \urlprefix  [0]{URL }%
\providecommand \Eprint [0]{\href }%
\providecommand \doibase [0]{https://doi.org/}%
\providecommand \selectlanguage [0]{\@gobble}%
\providecommand \bibinfo  [0]{\@secondoftwo}%
\providecommand \bibfield  [0]{\@secondoftwo}%
\providecommand \translation [1]{[#1]}%
\providecommand \BibitemOpen [0]{}%
\providecommand \bibitemStop [0]{}%
\providecommand \bibitemNoStop [0]{.\EOS\space}%
\providecommand \EOS [0]{\spacefactor3000\relax}%
\providecommand \BibitemShut  [1]{\csname bibitem#1\endcsname}%
\let\auto@bib@innerbib\@empty
\bibitem [{\citenamefont {DiTusa}(2006)}]{DiTusa2006}%
  \BibitemOpen
  \bibfield  {author} {\bibinfo {author} {\bibfnamefont {J.~F.}\ \bibnamefont
  {DiTusa}},\ }\bibfield  {title} {\bibinfo {title} {An integrated approach to
  physics seminars for students},\ }\href {https://doi.org/10.1119/1.2360995}
  {\bibfield  {journal} {\bibinfo  {journal} {Am. J. Phys.}\ }\textbf {\bibinfo
  {volume} {74}},\ \bibinfo {pages} {1045--1046} (\bibinfo {year}
  {2006})}\BibitemShut {NoStop}%
\bibitem [{\citenamefont {Dake}\ \emph {et~al.}(2018)\citenamefont {Dake},
  \citenamefont {Ribaudo},\ and\ \citenamefont {Day}}]{Dake2018}%
  \BibitemOpen
  \bibfield  {author} {\bibinfo {author} {\bibfnamefont {L.~S.}\ \bibnamefont
  {Dake}}, \bibinfo {author} {\bibfnamefont {J.}~\bibnamefont {Ribaudo}},\ and\
  \bibinfo {author} {\bibfnamefont {L.~H.}\ \bibnamefont {Day}},\ }\bibfield
  {title} {\bibinfo {title} {A multilevel seminar for physics majors: A good
  deal for everyone},\ }\href {https://doi.org/10.1119/1.5080584} {\bibfield
  {journal} {\bibinfo  {journal} {Phys. Teach.}\ }\textbf {\bibinfo {volume}
  {56}},\ \bibinfo {pages} {630--632} (\bibinfo {year} {2018})}\BibitemShut
  {NoStop}%
\bibitem [{\citenamefont {Cetnar}(2021)}]{Cetnar2021}%
  \BibitemOpen
  \bibfield  {author} {\bibinfo {author} {\bibfnamefont {A.~J.}\ \bibnamefont
  {Cetnar}},\ }\bibfield  {title} {\bibinfo {title} {Model for implementation
  of a modern journal club in medical physics residency programs},\ }\href
  {https://doi.org/10.1002/acm2.13250} {\bibfield  {journal} {\bibinfo
  {journal} {J. Appl. Clin. Med. Phys.}\ }\textbf {\bibinfo {volume} {22}},\
  \bibinfo {pages} {253--261} (\bibinfo {year} {2021})}\BibitemShut {NoStop}%
\bibitem [{\citenamefont {Santos}\ \emph {et~al.}(2021)\citenamefont {Santos},
  \citenamefont {Goto}, \citenamefont {Lu}, \citenamefont {Ho}, \citenamefont
  {Wang}, \citenamefont {On}, \citenamefont {Hashimoto},\ and\ \citenamefont
  {Young}}]{Santos2021}%
  \BibitemOpen
  \bibfield  {author} {\bibinfo {author} {\bibfnamefont {D.~J.~D.}\
  \bibnamefont {Santos}}, \bibinfo {author} {\bibfnamefont {T.}~\bibnamefont
  {Goto}}, \bibinfo {author} {\bibfnamefont {T.-Y.}\ \bibnamefont {Lu}},
  \bibinfo {author} {\bibfnamefont {S.~C.-C.}\ \bibnamefont {Ho}}, \bibinfo
  {author} {\bibfnamefont {T.-W.}\ \bibnamefont {Wang}}, \bibinfo {author}
  {\bibfnamefont {A.~Y.~L.}\ \bibnamefont {On}}, \bibinfo {author}
  {\bibfnamefont {T.}~\bibnamefont {Hashimoto}},\ and\ \bibinfo {author}
  {\bibfnamefont {S.~S.~C.}\ \bibnamefont {Young}},\ }\bibfield  {title}
  {\bibinfo {title} {Investigative study on preprint journal club as an
  effective method of teaching latest knowledge in astronomy},\ }\href
  {https://doi.org/10.1103/PhysRevPhysEducRes.17.010145} {\bibfield  {journal}
  {\bibinfo  {journal} {Phys. Rev. Phys. Educ. Res.}\ }\textbf {\bibinfo
  {volume} {17}},\ \bibinfo {pages} {010145} (\bibinfo {year}
  {2021})}\BibitemShut {NoStop}%
\bibitem [{\citenamefont {Corbo}\ \emph {et~al.}(2013)\citenamefont {Corbo},
  \citenamefont {Gandhi}, \citenamefont {Lee},\ and\ \citenamefont
  {Roth}}]{Corbo2013}%
  \BibitemOpen
  \bibfield  {author} {\bibinfo {author} {\bibfnamefont {J.}~\bibnamefont
  {Corbo}}, \bibinfo {author} {\bibfnamefont {P.}~\bibnamefont {Gandhi}},
  \bibinfo {author} {\bibfnamefont {G.}~\bibnamefont {Lee}},\ and\ \bibinfo
  {author} {\bibfnamefont {N.}~\bibnamefont {Roth}},\ }\bibfield  {title}
  {\bibinfo {title} {The Compass Project: Charting a new course in physics
  education},\ }\bibfield  {journal} {\bibinfo  {journal} {Phys. Today}\ }\href
  {https://doi.org/10.1063/PT.4.0003} {10.1063/PT.4.0003} (\bibinfo {year}
  {2013})\BibitemShut {NoStop}%
\bibitem [{\citenamefont {Sabella}\ \emph {et~al.}(2017)\citenamefont
  {Sabella}, \citenamefont {Mardis}, \citenamefont {Sanders},\ and\
  \citenamefont {Little}}]{Sabella2017}%
  \BibitemOpen
  \bibfield  {author} {\bibinfo {author} {\bibfnamefont {M.~S.}\ \bibnamefont
  {Sabella}}, \bibinfo {author} {\bibfnamefont {K.~L.}\ \bibnamefont {Mardis}},
  \bibinfo {author} {\bibfnamefont {N.}~\bibnamefont {Sanders}},\ and\ \bibinfo
  {author} {\bibfnamefont {A.}~\bibnamefont {Little}},\ }\bibfield  {title}
  {\bibinfo {title} {The Chi-Sci Scholars Program: Developing community and
  challenging racially inequitable measures of success at a minority-serving
  institution on chicago’s southside},\ }\href
  {https://doi.org/10.1119/1.4999730} {\bibfield  {journal} {\bibinfo
  {journal} {Phys. Teach.}\ }\textbf {\bibinfo {volume} {55}},\ \bibinfo
  {pages} {350} (\bibinfo {year} {2017})}\BibitemShut {NoStop}%
\bibitem [{\citenamefont {Quan}\ \emph {et~al.}(2019)\citenamefont {Quan},
  \citenamefont {Gutman}, \citenamefont {Corbo}, \citenamefont {Pollard},\ and\
  \citenamefont {Turpen}}]{Quan2019}%
  \BibitemOpen
  \bibfield  {author} {\bibinfo {author} {\bibfnamefont {G.}~\bibnamefont
  {Quan}}, \bibinfo {author} {\bibfnamefont {B.}~\bibnamefont {Gutman}},
  \bibinfo {author} {\bibfnamefont {J.}~\bibnamefont {Corbo}}, \bibinfo
  {author} {\bibfnamefont {B.}~\bibnamefont {Pollard}},\ and\ \bibinfo {author}
  {\bibfnamefont {C.}~\bibnamefont {Turpen}},\ }\bibfield  {title} {\bibinfo
  {title} {The Access Network: Cultivating equity and student leadership in
  STEM},\ }\bibfield  {journal} {\bibinfo  {journal} {Phys. Educ. Res. Conf.}\
  }\href {https://doi.org/10.1119/perc.2019.pr.Quan}
  {10.1119/perc.2019.pr.Quan} (\bibinfo {year} {2019})\BibitemShut {NoStop}%
\bibitem [{\citenamefont {Brown}\ and\ \citenamefont
  {Gonzales}(2021)}]{Brown2021}%
  \BibitemOpen
  \bibfield  {author} {\bibinfo {author} {\bibfnamefont {C.~D.}\ \bibnamefont
  {Brown}}\ and\ \bibinfo {author} {\bibfnamefont {E.}~\bibnamefont
  {Gonzales}},\ }\bibfield  {title} {\bibinfo {title} {Excellence and power in
  the Black physics community},\ }\href
  {https://doi.org/10.1038/s41567-020-01140-9} {\bibfield  {journal} {\bibinfo
  {journal} {Nat. Phys.}\ }\textbf {\bibinfo {volume} {17}},\ \bibinfo {pages}
  {3} (\bibinfo {year} {2021})}\BibitemShut {NoStop}%
\bibitem [{\citenamefont {Rethman}\ \emph {et~al.}(2021)\citenamefont
  {Rethman}, \citenamefont {Perry}, \citenamefont {Donaldson}, \citenamefont
  {Choi},\ and\ \citenamefont {Erukhimova}}]{Rethman2021}%
  \BibitemOpen
  \bibfield  {author} {\bibinfo {author} {\bibfnamefont {C.}~\bibnamefont
  {Rethman}}, \bibinfo {author} {\bibfnamefont {J.}~\bibnamefont {Perry}},
  \bibinfo {author} {\bibfnamefont {J.~P.}\ \bibnamefont {Donaldson}}, \bibinfo
  {author} {\bibfnamefont {D.}~\bibnamefont {Choi}},\ and\ \bibinfo {author}
  {\bibfnamefont {T.}~\bibnamefont {Erukhimova}},\ }\bibfield  {title}
  {\bibinfo {title} {Impact of informal physics programs on university student
  development: Creating a physicist},\ }\href
  {https://doi.org/10.1103/PhysRevPhysEducRes.17.020110} {\bibfield  {journal}
  {\bibinfo  {journal} {Phys. Rev. Phys. Educ. Res.}\ }\textbf {\bibinfo
  {volume} {17}},\ \bibinfo {pages} {020110} (\bibinfo {year}
  {2021})}\BibitemShut {NoStop}%
\bibitem [{\citenamefont {Mermin}(2000)}]{Mermin2000}%
  \BibitemOpen
  \bibfield  {author} {\bibinfo {author} {\bibfnamefont {N.~D.}\ \bibnamefont
  {Mermin}},\ }\bibfield  {title} {\bibinfo {title} {The contemplation of
  quantum computation},\ }\href {https://doi.org/10.1063/1.1292466} {\bibfield
  {journal} {\bibinfo  {journal} {Phys. Today}\ }\textbf {\bibinfo {volume}
  {53}},\ \bibinfo {pages} {11} (\bibinfo {year} {2000})}\BibitemShut {NoStop}%
\bibitem [{\citenamefont {Wood}(2020)}]{Wood2020}%
  \BibitemOpen
  \bibfield  {author} {\bibinfo {author} {\bibfnamefont {C.}~\bibnamefont
  {Wood}},\ }\bibfield  {title} {\bibinfo {title} {The cartoon picture of
  magnets that has transformed science},\ }\href
  {https://www.quantamagazine.org/the-cartoon-picture-of-magnets-that-has-transformed-science-20200624/}
  {\bibfield  {journal} {\bibinfo  {journal} {Quanta Magazine}\ } (\bibinfo
  {year} {June 24, 2020})}\BibitemShut {NoStop}%
\bibitem [{\citenamefont {Weatherall}(2017)}]{Weatherall2017}%
  \BibitemOpen
  \bibfield  {author} {\bibinfo {author} {\bibfnamefont {J.~O.}\ \bibnamefont
  {Weatherall}},\ }\bibfield  {title} {\bibinfo {title} {Is quantum theory
  about reality or what we know?},\ }\href
  {http://nautil.us/blog/-is-quantum-theory-about-reality-or-what-we-know}
  {\bibfield  {journal} {\bibinfo  {journal} {Nautilus}\ } (\bibinfo {year}
  {December 4, 2017})}\BibitemShut {NoStop}%
\bibitem [{\citenamefont {Redish}\ and\ \citenamefont
  {Steinberg}(1999)}]{Redish1999}%
  \BibitemOpen
  \bibfield  {author} {\bibinfo {author} {\bibfnamefont {E.~F.}\ \bibnamefont
  {Redish}}\ and\ \bibinfo {author} {\bibfnamefont {R.~N.}\ \bibnamefont
  {Steinberg}},\ }\bibfield  {title} {\bibinfo {title} {Teaching physics:
  Figuring out what works},\ }\href {https://doi.org/10.1063/1.882568}
  {\bibfield  {journal} {\bibinfo  {journal} {Phys. Today}\ }\textbf {\bibinfo
  {volume} {52}},\ \bibinfo {pages} {24} (\bibinfo {year} {1999})}\BibitemShut
  {NoStop}%
\bibitem [{\citenamefont
  {Castelvecchi}(2019{\natexlab{a}})}]{Castelvecchi2019}%
  \BibitemOpen
  \bibfield  {author} {\bibinfo {author} {\bibfnamefont {D.}~\bibnamefont
  {Castelvecchi}},\ }\bibfield  {title} {\bibinfo {title} {First hint of
  near-room-temperature superconductor tantalizes physicists},\ }\href
  {https://doi.org/10.1038/d41586-018-07831-x} {\bibfield  {journal} {\bibinfo
  {journal} {Nature}\ }\textbf {\bibinfo {volume} {565}},\ \bibinfo {pages}
  {12} (\bibinfo {year} {2019}{\natexlab{a}})}\BibitemShut {NoStop}%
\bibitem [{\citenamefont {Gray}\ and\ \citenamefont
  {Mansouli\'e}(2018)}]{Gray2018}%
  \BibitemOpen
  \bibfield  {author} {\bibinfo {author} {\bibfnamefont {H.}~\bibnamefont
  {Gray}}\ and\ \bibinfo {author} {\bibfnamefont {B.}~\bibnamefont
  {Mansouli\'e}},\ }\bibfield  {title} {\bibinfo {title} {The Higgs boson: the
  hunt, the discovery, the study and some future perspectives},\ }\href
  {https://atlas.cern/updates/feature/higgs-boson} {\bibfield  {journal}
  {\bibinfo  {journal} {ATLAS Updates}\ } (\bibinfo {year} {2018})}\BibitemShut
  {NoStop}%
\bibitem [{\citenamefont {Levi}(2002)}]{Levi2002}%
  \BibitemOpen
  \bibfield  {author} {\bibinfo {author} {\bibfnamefont {B.~G.}\ \bibnamefont
  {Levi}},\ }\bibfield  {title} {\bibinfo {title} {Investigation finds that one
  Lucent physicist engaged in scientific misconduct},\ }\href
  {https://doi.org/10.1063/1.1534995} {\bibfield  {journal} {\bibinfo
  {journal} {Phys. Today}\ }\textbf {\bibinfo {volume} {55}},\ \bibinfo {pages}
  {15} (\bibinfo {year} {2002})}\BibitemShut {NoStop}%
\bibitem [{\citenamefont {Couzin}(2006)}]{Couzin2006}%
  \BibitemOpen
  \bibfield  {author} {\bibinfo {author} {\bibfnamefont {J.}~\bibnamefont
  {Couzin}},\ }\bibfield  {title} {\bibinfo {title} {New optics strategies cut
  through diffraction barrier},\ }\href
  {https://doi.org/10.1126/science.313.5788.748a} {\bibfield  {journal}
  {\bibinfo  {journal} {Science}\ }\textbf {\bibinfo {volume} {313}},\ \bibinfo
  {pages} {748} (\bibinfo {year} {2006})}\BibitemShut {NoStop}%
\bibitem [{\citenamefont {Betzig}\ \emph {et~al.}(2006)\citenamefont {Betzig},
  \citenamefont {Patterson}, \citenamefont {Sougrat}, \citenamefont
  {Lindwasser}, \citenamefont {Olenych}, \citenamefont {Bonifacino},
  \citenamefont {Davidson}, \citenamefont {Lippincott-Schwartz},\ and\
  \citenamefont {Hess}}]{Betzig2006}%
  \BibitemOpen
  \bibfield  {author} {\bibinfo {author} {\bibfnamefont {E.}~\bibnamefont
  {Betzig}}, \bibinfo {author} {\bibfnamefont {G.~H.}\ \bibnamefont
  {Patterson}}, \bibinfo {author} {\bibfnamefont {R.}~\bibnamefont {Sougrat}},
  \bibinfo {author} {\bibfnamefont {O.~W.}\ \bibnamefont {Lindwasser}},
  \bibinfo {author} {\bibfnamefont {S.}~\bibnamefont {Olenych}}, \bibinfo
  {author} {\bibfnamefont {J.~S.}\ \bibnamefont {Bonifacino}}, \bibinfo
  {author} {\bibfnamefont {M.~W.}\ \bibnamefont {Davidson}}, \bibinfo {author}
  {\bibfnamefont {J.}~\bibnamefont {Lippincott-Schwartz}},\ and\ \bibinfo
  {author} {\bibfnamefont {H.~F.}\ \bibnamefont {Hess}},\ }\bibfield  {title}
  {\bibinfo {title} {Imaging intracellular fluorescent proteins at nanometer
  resolution},\ }\href {https://doi.org/10.1126/science.1127344} {\bibfield
  {journal} {\bibinfo  {journal} {Science}\ }\textbf {\bibinfo {volume}
  {313}},\ \bibinfo {pages} {1642} (\bibinfo {year} {2006})}\BibitemShut
  {NoStop}%
\bibitem [{\citenamefont {{SLAC National Accelerator Laboratory}}()}]{slac}%
  \BibitemOpen
  \bibfield  {author} {\bibinfo {author} {\bibnamefont {{SLAC National
  Accelerator Laboratory}}},\ }\href@noop {} {\bibinfo {title} {48th {SLAC}
  summer institute contribution list}},\ \bibinfo {howpublished}
  {\url{https://indico.slac.stanford.edu/event/326/contributions/}},\ \bibinfo
  {note} {accessed: 2022-08-31}\BibitemShut {NoStop}%
\bibitem [{\citenamefont {Merali}(2020)}]{Merali2020}%
  \BibitemOpen
  \bibfield  {author} {\bibinfo {author} {\bibfnamefont {Z.}~\bibnamefont
  {Merali}},\ }\bibfield  {title} {\bibinfo {title} {This twist on
  Schrödinger’s cat paradox has major implications for quantum theory},\
  }\href
  {https://www.scientificamerican.com/article/this-twist-on-schroedingers-cat-paradox-has-major-implications-for-quantum-theory/}
  {\bibfield  {journal} {\bibinfo  {journal} {Sci. Am.}\ } (\bibinfo {year}
  {August 17, 2020})}\BibitemShut {NoStop}%
\bibitem [{\citenamefont
  {Castelvecchi}(2019{\natexlab{b}})}]{Castelvecchi2019a}%
  \BibitemOpen
  \bibfield  {author} {\bibinfo {author} {\bibfnamefont {D.}~\bibnamefont
  {Castelvecchi}},\ }\bibfield  {title} {\bibinfo {title} {Black hole imaged
  for first time},\ }\href {https://doi.org/10.1038/d41586-019-01155-0}
  {\bibfield  {journal} {\bibinfo  {journal} {Nature}\ }\textbf {\bibinfo
  {volume} {568}},\ \bibinfo {pages} {284} (\bibinfo {year}
  {2019}{\natexlab{b}})}\BibitemShut {NoStop}%
\bibitem [{\citenamefont {Schirber}(2018)}]{Schirber2018}%
  \BibitemOpen
  \bibfield  {author} {\bibinfo {author} {\bibfnamefont {M.}~\bibnamefont
  {Schirber}},\ }\bibfield  {title} {\bibinfo {title} {Pinning down superheavy
  masses},\ }\href {https://physics.aps.org/articles/v11/s137} {\bibfield
  {journal} {\bibinfo  {journal} {Physics}\ }\textbf {\bibinfo {volume} {11}},\
  \bibinfo {pages} {s137} (\bibinfo {year} {2018})}\BibitemShut {NoStop}%
\bibitem [{\citenamefont {Paul}\ \emph {et~al.}(2018)\citenamefont {Paul},
  \citenamefont {Webb}, \citenamefont {Chessey},\ and\ \citenamefont
  {Lucas}}]{Paul2018}%
  \BibitemOpen
  \bibfield  {author} {\bibinfo {author} {\bibfnamefont {C.}~\bibnamefont
  {Paul}}, \bibinfo {author} {\bibfnamefont {D.~J.}\ \bibnamefont {Webb}},
  \bibinfo {author} {\bibfnamefont {M.~K.}\ \bibnamefont {Chessey}},\ and\
  \bibinfo {author} {\bibfnamefont {J.}~\bibnamefont {Lucas}},\ }\bibfield
  {title} {\bibinfo {title} {Pondering zeros: Uncovering hidden inequities
  within a decade of grades},\ }\bibfield  {journal} {\bibinfo  {journal}
  {Phys. Educ. Res. Conf.}\ }\href {https://doi.org/10.1119/perc.2018.pr.Paul}
  {10.1119/perc.2018.pr.Paul} (\bibinfo {year} {2018})\BibitemShut {NoStop}%
\bibitem [{\citenamefont {Guskey}(2013)}]{Guskey2013}%
  \BibitemOpen
  \bibfield  {author} {\bibinfo {author} {\bibfnamefont {T.~R.}\ \bibnamefont
  {Guskey}},\ }\bibfield  {title} {\bibinfo {title} {The case against
  percentage grades},\ }\href {https://uknowledge.uky.edu/edp_facpub/22/}
  {\bibfield  {journal} {\bibinfo  {journal} {Educ. Leadership}\ }\textbf
  {\bibinfo {volume} {71}},\ \bibinfo {pages} {68} (\bibinfo {year}
  {2013})}\BibitemShut {NoStop}%
\bibitem [{\citenamefont {Savitsky}(2020)}]{Savitsky2020}%
  \BibitemOpen
  \bibfield  {author} {\bibinfo {author} {\bibfnamefont {Z.}~\bibnamefont
  {Savitsky}},\ }\bibfield  {title} {\bibinfo {title} {The next big thing: the
  use of graph neural networks to discover particles},\ }\href
  {https://news.fnal.gov/2020/09/the-next-big-thing-the-use-of-graph-neural-networks-to-discover-particles/}
  {\bibfield  {journal} {\bibinfo  {journal} {Fermilab Newsroom}\ } (\bibinfo
  {year} {2020})}\BibitemShut {NoStop}%
\bibitem [{\citenamefont {Aurisano}\ \emph {et~al.}(2016)\citenamefont
  {Aurisano}, \citenamefont {Radovic}, \citenamefont {Rocco}, \citenamefont
  {Himmel}, \citenamefont {Messier}, \citenamefont {Niner}, \citenamefont
  {Pawloski}, \citenamefont {Psihas}, \citenamefont {Sousa},\ and\
  \citenamefont {Vahle}}]{Aurisano2016}%
  \BibitemOpen
  \bibfield  {author} {\bibinfo {author} {\bibfnamefont {A.}~\bibnamefont
  {Aurisano}}, \bibinfo {author} {\bibfnamefont {A.}~\bibnamefont {Radovic}},
  \bibinfo {author} {\bibfnamefont {D.}~\bibnamefont {Rocco}}, \bibinfo
  {author} {\bibfnamefont {A.}~\bibnamefont {Himmel}}, \bibinfo {author}
  {\bibfnamefont {M.}~\bibnamefont {Messier}}, \bibinfo {author} {\bibfnamefont
  {E.}~\bibnamefont {Niner}}, \bibinfo {author} {\bibfnamefont
  {G.}~\bibnamefont {Pawloski}}, \bibinfo {author} {\bibfnamefont
  {F.}~\bibnamefont {Psihas}}, \bibinfo {author} {\bibfnamefont
  {A.}~\bibnamefont {Sousa}},\ and\ \bibinfo {author} {\bibfnamefont
  {P.}~\bibnamefont {Vahle}},\ }\bibfield  {title} {\bibinfo {title} {A
  convolutional neural network neutrino event classifier},\ }\href
  {https://doi.org/10.1088/1748-0221/11/09/p09001} {\bibfield  {journal}
  {\bibinfo  {journal} {J. Instrum.}\ }\textbf {\bibinfo {volume} {11}}\bibinfo
   {number} { (09)},\ \bibinfo {pages} {P09001}}\BibitemShut {NoStop}%
\bibitem [{\citenamefont {Michelson}\ and\ \citenamefont
  {Morley}(1887)}]{Michelson1887}%
  \BibitemOpen
\bibfield  {number} {  }\bibfield  {author} {\bibinfo {author} {\bibfnamefont
  {A.}~\bibnamefont {Michelson}}\ and\ \bibinfo {author} {\bibfnamefont
  {E.}~\bibnamefont {Morley}},\ }\bibfield  {title} {\bibinfo {title} {On the
  relative motion of the Earth and the luminiferous ether},\ }\href
  {https://doi.org/10.2475/ajs.s3-34.203.333} {\bibfield  {journal} {\bibinfo
  {journal} {Am. J. Sci.}\ }\textbf {\bibinfo {volume} {34}},\ \bibinfo {pages}
  {334} (\bibinfo {year} {1887})}\BibitemShut {NoStop}%
\bibitem [{\citenamefont {{NASA Earth Observatory}}()}]{climate1}%
  \BibitemOpen
  \bibfield  {author} {\bibinfo {author} {\bibnamefont {{NASA Earth
  Observatory}}},\ }\href@noop {} {\bibinfo {title} {Climate and {E}arth’s
  energy budget}},\ \bibinfo {howpublished}
  {\url{https://earthobservatory.nasa.gov/features/EnergyBalance}},\ \bibinfo
  {note} {accessed: 2022-08-31}\BibitemShut {NoStop}%
\bibitem [{\citenamefont {{Lawrence Livermore National Laboratory}}()}]{nif}%
  \BibitemOpen
  \bibfield  {author} {\bibinfo {author} {\bibnamefont {{Lawrence Livermore
  National Laboratory}}},\ }\href@noop {} {\bibinfo {title} {How {NIF}
  works}},\ \bibinfo {howpublished}
  {\url{https://lasers.llnl.gov/about/how-nif-works}},\ \bibinfo {note}
  {accessed: 2022-08-31}\BibitemShut {NoStop}%
\end{thebibliography}
%

\end{document}